\documentclass[11pt]{article}
\usepackage{amsmath,amssymb,graphicx}
\usepackage{amsthm}
\usepackage{euscript}
\usepackage{eufrak}

\newtheorem{thm}{Theorem}

\newcommand{\pa}{\partial}

\newcommand{\const}{\textrm{const}}

\newcommand{\supp}{ \mathrm{ supp  }}
\newcommand{\hs}{ \hspace{1cm}}

\newcommand{\Vol}{\textrm{Vol}}

\newcommand{\B}{\Big}

\newcommand{\crit}{\textrm{crit}}

\newcommand{\be}{\begin{equation}} 
\newcommand{\ee}{\end{equation}} 

\newcommand{\al}{\alpha}
\newcommand{\De}{\Delta}
\newcommand{\de}{\delta}
\newcommand{\ga}{\gamma}
\newcommand{\Ga}{\Gamma}
\newcommand{\ka}{\kappa}

\newcommand{\Om}{\Omega}
\newcommand{\om}{\omega}

\newcommand{\ep}{\epsilon}

\newcommand{\vep}{\varepsilon}
\newcommand{\bpsi}{\bar \psi}
\newcommand{\bPsi}{\bar \Psi}

\newcommand{\cO}{{\cal O}}

\newcommand{\cH}{{\cal H}}

\newcommand{\cR}{{\cal R}}

\newcommand{\cW}{{\cal W}}
\newcommand{\cQ}{{\cal Q}}
\newcommand{\cL}{{\cal L}}

\newcommand{\cZ}{{\cal Z}}

\newcommand{\bbZ}{{\mathbb{Z}}}

\newcommand{\bbT}{{\mathbb{T}}}

\newcommand{\cA}{ \EuScript{A} }
\newcommand{\fD}{\EuFrak{D}} 
\newcommand{\fS}{\EuFrak{S}}
\newcommand{\fZ} {\EuFrak{Z}} 

\newcommand{\sZ}{\mathsf{Z}}

\newcommand{\sx}{\mathsf{x}}
\newcommand{\sN}{\mathsf{N}}

\newcommand{\bom}{ \mathbf{\Om} }
\newcommand{\sq}{\square} 
\newcommand{\bW}{\bar W} 

\begin{document}

\title{Stability for  QED in d=3 : an overview }
\author{ 
J. Dimock
\thanks{dimock@buffalo.edu. }
\thanks{Expanded version of talk given at ICMP 2021, Geneva}\\
Dept. of Mathematics \\
SUNY at Buffalo \\
Buffalo, NY 14260 }
\maketitle

\begin{abstract}
We report on a  result on quantum electrodynamics on a three dimensional Euclidean spacetime.  
The model is formulated  on a toroidal lattice with unit volume  and  variable lattice spacing.  The result is
that  the renormalized partition function  is bounded above and below uniformly in the lattice spacing.   This is a first step 
toward  showing that  the partition function   and  correlation functions have  limits  as the  lattice spacing goes to zero.
\end{abstract}

\section{Introduction}

The topic is quantum electrodynamics (QED) on a three dimensional Euclidean spacetime, one short of the physical dimension. It is   a quantum field theory 
for relativistic electrons and photons.    The model is defined initially in finite volume and with a short distance regularization.
A  basic problem in the construction of the model  is to remove the regularization and take the infinite volume limit.   In this note we report on 
some recent progress  toward the first of these problems.

The model is formulated on  a 3-dimensional toroidal lattice.   
These will have the general form 
\be
\bbT^{-N}_{N'}  =     ( L^{-N} \bbZ/L^{N'}  \bbZ)^3  
\ee
 where $L$ is a fixed large positive number.   We start with  
  $\bbT^{-N}_0 $  which has lattice spacing $\ep \equiv  L^{-N }$ and  unit volume.  
  The theory is defined by the action 
\be
S(e, \cA,   \psi)   =    \frac12   \| d \cA  \|^2 +  <  \bar \psi,   (\fD _e (\cA)   +  \bar   m) \psi >    +   m^N  <  \bar \psi,   \psi > 
+   \vep^N   
\ee   
Here 
 $ \cA(b)  $ is an  abelian gauge field   defined  on bonds $b$ (nearest neighbor pairs)  in the lattice 
and 
$ d \cA(p) = \ep^{-1} \sum_{b \in \pa p} A(b) $     is the field strength defined on plaquettes (squares) in the lattice. 
The notation is   $\| d \cA ]|^2 = \sum_p  \ep^3 |d \cA (p) |^2$.  
The 
$ \bpsi_{\al} (x), \psi_{\al}(x)   $  are  fermi fields indexed by lattice points and spin indices  and  $<\bar \psi, \psi> =  \sum_{x, \al} \ep^3   \bar \psi_{\al}(x) \psi_{\al} (x) $.  The
fermi fields are anti-commuting     generators of  a Grassmann algebra.
The 
$\fD_e  (\cA) $  is the  Wilson form of the covariant  lattice   Dirac operator with charge $e$ defined by  
\be
\fD _e(\cA) = \ga \cdot \nabla_{e\cA}  - \frac12 \ep \De_{e\cA}  
\ee 
Here   $\{\ga_{\mu}, \ga_{\nu} \} = 2 \de_{\mu \nu}$ are Dirac matrices,   the  forward covariant derivative is
\be
  (\pa_{e\cA, \mu} \psi)(x)  = \ep^{-1} \B( e^{ i\ep e A( x, x+ \ep e_{\mu}) }\psi(x+ \ep e_{\mu} ) - \psi(x)\B), 
\ee 
$\nabla_{e\cA, \mu}$ is a symmetric version of this,  and $\De_{e \cA} $ is the associated Laplacian. 
The  $\bar m $ is a background mass and  
$ m^N, \vep^N$  are respectively mass and energy density counterterms which will be chosen to depend on the lattice spacing, thereby renormalizing these
quantities.  The model is super-renormalizable and no renormalization is required for charge or field strength.

A basic item of interest is   the partition function which is the exponential of the action integrated over all fields
\be 
 \sZ_N (e) =  \int    \exp (  -  S(e, \cA,   \psi)    ) \      D  \psi      \ D  \cA    
  \ee
Here  $D \cA = \prod_b d ( \cA(b) ) $  and $D \psi$ stands for   Grassmann integral over all fermi fields $\bpsi, \psi$, the projection onto the maximal element.     
Actually the integral over the gauge fields does not converge  because the integrand is invariant under gauge transformations.  One must fix a
gauge, that is pick a representative for  each gauge orbit and integrate over them.  Once this choice is made the partition function $ \sZ_N (e)$ is well-defined.   The
main  result is a bound on the  relative partition 
function uniform in the lattice  spacing:  

\begin{thm} \label{one} (Ultraviolet Stability) Assume  $\bar m>0$ and $e$ is  sufficiently small.  Then one  can  fix the gauge and choose  
counterterms  $m^N, \vep^N$ so  that for all $N$
\[ K_- \leq \frac{\sZ_N (e) }{\sZ_N (0)}   \leq K_+
\]
for some constants $K_{\pm}$.
\end{thm} 
\bigskip

The proof can be found in the series of papers \cite{Dim15b},\cite{Dim20a},\cite{Dim20b}  (see also \cite{Dim13a},\cite{Dim13b},\cite{Dim14}).    It uses a block averaging  renormalization group technique due to  Balaban, 
a rigorous version of earlier ideas of Wilson.   The method  
was introduced 
by Balaban  \cite{Bal82a}-\cite{Bal83b}  for scalar electrodynamics in d=3 .  It was further developed by    Balaban, Imbrie, and Jaffe \cite{BIJ85},\cite{BIJ88},\cite{Imb86} 
  for the abelian Higgs model in d=3 and by  Balaban  \cite{Bal84a}-\cite{Bal88b} for  Yang Mills  in d=3,4 . 
There is also   relevant  work on block averaged  Dirac  Green's functions by Balaban, O'Carroll, and Shor \cite{BOS89},\cite{BOS91}.

The proof should allow one to add local  gauge invariant  perturbations to the action like  $<dA, J>$ or $< \bpsi \ga_{\mu} \psi, j_{\mu} >$.   Then the partition function $ \sZ_N (e) $
becomes a generating function $ \sZ_N (e,J,j) $,  derivatives  of $\log  \sZ_N (e,J,j)$ in $J,j$  at $J=0,j=0$  give correlation functions.  Uniform bounds on the generating function
and analyticity in $J,j$ would give uniform bounds on correlation functions. 
The continuum limit $N \to \infty$ for the partition function, generating function,  and  correlation functions should also be feasible.  The  method works  equally well for any finite volume, not just
unit volume.   But the infinite volume limit would  pose new difficulties. 

In the remainder of the paper we  explain some of the ideas that go into the proof.  Before proceeding   it is convenient  to  first reformulate the problem by   scaling  up to a unit lattice.   If  $\Psi_0,  A_0$ are  
 fields on the large  unit lattice   $\bbT^0_{N} $, 
 we define a new action by   $ S_0 (e_0, \Psi_0, A_0)  = S(e, \Psi_{0,L^{-N} } , A_{0,L^{-N}}) $
 where  
  $\Psi_{0,L^{-N} } (x) = L^{N} \Psi_0( L^Nx)$ and $A_{0,L^{-N}}(b) =L^{N/2} A_0(L^Nb) $ are functions on the original  lattice $\bbT^{-N}_0$.
 We find that
\be \label{picky} 
S_0(e_0, A_0,   \Psi_0)   =    \frac12   \| d A_0  \|^2 +  <  \bar \Psi_0,   (\fD _{e_0 } (A_0 )   +  \bar   m_0) \Psi_0 >    +   m_0  <  \bar \Psi_0,   \Psi_0 > 
+     \vep_0  | \bbT^0_N|  
\ee   
Here  norms, inner products, and operators  are now  on the unit lattice, and   the parameters have  become tiny
\be
e_0 = e_0^N = L^{-N/2} e \hs    \bar m_0 = \bar m_0^N = L^{-N} \bar m 
\ee
as well as $  m_0 =  m_0^N = L^{-N} \bar m^N  $ and $\vep_0 = \vep_0^N = L^{-3N} \vep^N$.   Our   $N \to \infty$ problem now has become   an infinite volume problem with scaled
parameters.

\section{Renormalization group transformations}

The difficulty with the problem   is that we have an unbounded number of variables.   The renormalization group method consists of integrating out  a few variables at 
a time,  and keeping careful track of the effective actions at each stage.   (The transformations  on the actions  are not invertible and  do not actually form a group. 
Nevertheless the terminology "renormalization group" is ubiquitous.)  
 
In the block averaging method     one integrates out all fields  which have a fixed average over  $L$-cubes.
Let $y$ be a point in the $L$-lattice    $ \bbT^1_N $ and let $B(y)$ be the $L$-cube  in the unit lattice $\bbT_N^0$ centered on $y$. 
For fermion fields $\Psi_0$  the averaging operator in the presence of the gauge field $A_0$ is
    \be    \label{dice1}
(Q(A_0)\Psi_0 )(y)    =   L^{-3}   \sum_{x \in B(y)} e^{ ie_0 A_0(\Ga( y,x))   } \Psi_0  (x) 
 \ee
Here  $A_0(\Ga) =  \sum_{b \in \Ga}A_0(b) $   and     $\Ga(y,x) $ is a  path from $x$ to $ y$. We  use it to  parallel translate to the center of the cube before averaging, so 
that gauge covariance is  preserved. 
For the gauge field let $(y, y+ Le_{\mu})$ be a bond in the $L$-lattice, and  define  the averaged gauge field on this bond to be
\be   
(\cQ A_0) (y,  y + L e_{\mu} )  
=  \sum_{x \in B(y)  } L^{-4}    A_0(  \Ga_{x,  x +  L e_{\mu} } )  
\ee
This also preserves gauge covariance. 

With these definitions the renormalization group transformation sends the initial density $\rho_0  = e^{-S_0}  $ to a new density $\rho_1$
defined for fields $A_1,\Psi_1 $ defined on the $L$-lattice $\bbT^1_N$  by
\be   
\begin{split}     \label{pollypolly} 
\tilde \rho_1(A_1,  \Psi_1  )  & =  \sN_0
   \int  DA_0   D \Psi_0  \  \de ( A_1 - \cQ A_0)  \   \de_{\sx}  (A_0)     \\
   &  \exp\B( - b/L <  \bPsi_1 -  Q(- A_0 ) \bPsi_0,  \Psi_{1} -  Q(A_0 )  \Psi_0  > \B)    \rho_{0} (A_0,  \Psi_0)     \\
\end{split}    
\ee 
Here
 $b>0 $ is a constant and $\sN_0$ normalizes the fermion integral.  
We have avoided using a delta function for the fermion fields   (which would not work for Grassmann elements) 
and instead used a Gaussian factor.    We have also begun to fix the gauge by taking a certain maximal tree in each $L$-cube,
 and    inserting a delta function $ \de_{\sx}  (A_0) $ that sets $A_0 = 0$ on the bonds in each tree.  With this  axial gauge, and the fixed averages,   the   integral over the remaining  fields converges  and  $\tilde \rho_1$ is well-defined.

  After the block averaging we scale back down defining for  $A_1, \Psi_1$ on  the  unit lattice $\bbT^0_{N-1} $ 
 \be \label{boots1}
  \rho_1(A_1, \Psi_1) = \tilde \rho_1(A_{1,L}, \Psi_{1,L}  ) 
 \ee
 where  $
  A_{1,L} (b)  = L^{-\frac12} A_1(b/L)$ and 
 $ \Psi_{1,L}(x)  = L^{-1} \Psi_1(x/L) $
  are fields on  $\bbT^1_N $.

We would like the density $\rho_1$  to give the same partition function  as  $\rho_0$.
Formally this is true since  it   satisfies, up to an overall constant,  
\be \label{sunrise} 
\int  \de_{\sx} (A_1) \rho_1(A_1,  \Psi_1  ) DA_1 D \Psi_1 
= \int   \de_{\sx} (\cQ A_0 )  \de_{\sx} (A_0)  \rho_0(A_0,  \Psi_0  )DA_0   D \Psi_0 
\ee
However  these integrals do not converge  since we do not as yet have a complete gauge fixing. 
 Nevertheless one can repeat   the transformation   generating a sequence $\rho_0, \rho_1, \rho_2, \dots$.  After $k$ steps this yields a density $\rho_k(A_k, \Psi_k) $
 where the fields are defined on the unit lattice $ \bbT^0_{N-k} $ with shrunken volume.   We continue to iterate for $K$ steps with $K$ close to $N$ .
  Then we have a density $\rho_K $  on the lattice    $\bbT^0_{N-K}$ with bounded volume. 
Instead of 
  (\ref{sunrise}) we get    something like  
\be  \label{sunset} 
\int  \de_{\sx} (A_K) \rho_K(A_K,  \Psi_K   ) DA_K D \Psi_K    = 
 \int    \B(     \prod^K _{j=0} \de_{\sx} (\cQ ^j A_0 ) \B)   \rho_0 (A_0,  \Psi_0  ) DA_0 D \Psi_0 
\ee 
The right side  of (\ref{sunset})  does have a  more or less complete gauge fixing \cite{Dim20b}.  It is an axial gauge with a hierarchical structure,  and it provides the   
gauge fixed definition of the partition function  $Z_N(e)$.  Thus the identity (\ref{sunset}) shows that the partition function can be computed in principle   from  $\rho_K $ which has  a bounded number of variables. 
Thus the issue is   to keep control over  the sequence $\rho_0, \rho_1, \rho_2, \dots $.

\section{Bounded gauge fields} \label{split}

In analyzing the flow of the renormalization group it is useful to have bounds on the gauge field,  the fermion field is effectively already  bounded. 
We discuss how to arrange this in a simple case.    The actual implementation is rather more complicated, but the following gives the basic idea.

  In the initial density  we have a factor $ e^{-\frac12 \| dA_0 \|^2 } $ which is going to suppress large values 
of $dA_0 $.   To translate this fact into hard bounds on $dA_0$ we divide the torus into cubes of size $M= L^m$ larger than $L$.   Then we  introduce the characteristic functions for each $M$-cube $\sq$ by 
\be
 \chi_0 ( , \sq, A_0 )  = \chi\B( \supp_{p \in \sq } |dA_0(p)| \leq  p(e_0)  \B)
\ee
Here 
\be 
p(e_0) = ( - \log e_0 )^p 
\ee
 for some positive integer $p$.   Since $e_0$ is tiny  $p(e_0)$ is large,  but   not 
too large in the sense that $e_0p(e_0) $ is still very small.   Then with $\zeta_0 (\sq) = 1 -  \chi_0(\sq) $  we write
\be \label{tiger1}
\begin{split}
1 =&   \prod_{\square } \zeta_0( \sq) +  \chi_0( \sq) 
=    \sum_{\Om}  \prod_{\sq \subset  \Om^c } \zeta_0( \square) 
\prod_{\sq  \subset  \Om}  \chi_0 (\sq)  
\equiv    \sum_{\Om} \zeta_0( \Om^c)    \chi_0(\Om)  \\
\end{split}
\ee
where $\Om$ is an  arbitrary  union of $M$-cubes.  The   identity  could be  inserted  under the integral sign in (\ref{polly}) and then the sum  taken outside the integral.  For the term 
$\Om$  the  characteristic function  $\chi_0(\Om) $ enforces that   $|dA_0(p)| \leq  p(e_0) $ everywhere in $\Om$.  The characteristic function 
 $\zeta_0( \Om^c)  $ enforces that in each $M$-cube $\sq $ in $\Om^c$  there is at least one plaquette $p$ such that $|dA_0(p) | > p(e_0)$.
 It follow that for  $\sq \subset \Om^c$
 \be \label{swoosh} 
   \zeta_0 (\sq) e^{-\frac14 \| dA_0 \|^2_{\sq}   }  \leq   e^{ -\frac14 p(e_0)^2 } 
\ee
The quantity    $ e^{ -\frac14 p(e_0)^2 } $ is in fact a very small  number;  it satisfies     $ e^{ -\frac14 p(e_0)^2 }= \cO(e_0^n) $ for any positive integer $n$. 
The bound (\ref{swoosh}) gives that 
\be
 \zeta_0 (\Om^c)  )  e^{-\frac14 \| dA_0 \|^2_{\Om^c}   }  \leq   e^{ -\frac14 p(e_0)^2| \Om^c|_M } 
\ee
where $|\Om^c|_M$ is the number of $M$ cubes in $\Om^c$.
This is enough for the convergence of the sum over $\Om$ for we have 
\be
\sum_{\Om}  e^{ -\frac14 p(e_0)^2|\Om^c|_M  }  = \prod_{\sq} ( 1 +   e^{ -\frac14 p(e_0)^2 } ) \leq  \prod_{\sq} \exp (    e^{ -\frac14 p(e_0)^2}   )
 \leq   \exp \B(    e^{ -\frac14 p(e_0)^2 }  | \bbT^0_N |_M  \B) 
\ee
The last expression is   $\exp( \cO(e^n) ) \approx 1 $   since      $ e^{ -\frac14 p(e_0)^2 } = \cO ( L^{-nN/2} e^n )  $   beats $ | \bbT^0_N |_M  = L^{3N}  M^{-3}  $.

Tiny factors like $  e^{ -\frac14 p(e_0)^2 } $ also ensure  that contributions  from   large field regions $\Om^c$  are  negligible  relative to the contribution of  the
small  field region $\Om$.  They do not require renormalization  and can be bounded without further processing.   The  contribution of a small field region
does require careful analysis and renormalization as we will explain.

\section{The first step}

\label{four}

Starting with $\rho_0 =e^{-S_0}$ defined in (\ref{picky}), 
the density after the first step defined in (\ref{pollypolly}) has the form 
\be       \label{polly2} 
\tilde   \rho_1(A_1,  \Psi_1  )   =     \int      DA_0  \   \de ( A_1 - \cQ A_0)  \de_{\sx} (A_0)  \exp\B( -    \frac12   \| d A_0  \|^2 \B)  \rho_f(A_0,\Psi_1) 
\ee
where the fermion integral is
\be
\begin{split}
 \rho_f(A_0,\Psi_1)   = & \sN_0
 \int     D \Psi_0   \exp\B( - bL^{-1}   <   \bPsi_1 -  Q(- A_0 ) \bPsi_0,  \Psi_{1} -  Q(A_0 )  \Psi_k >\\
 & - <  \bar \Psi_0,   (\fD_{e_0}  (A_0)  + \bar    m_0 )\Psi_0>  - m_0  <  \bar \Psi_0,   \Psi_0 > 
-     \vep_0  | \bbT^0_N|    \B)   \\
 \end{split} 
\ee

Our strategy  is to make some large/small field splits, and then in the small field region write the new density  as the exponential of an effective action.
In the following for simplicity we only discuss the most important case where the small field region is the full torus.  Furthermore we do not always
explicitly record the characteristic functions bounding the fields.

First for the gauge field  we define  $A^{\min, \sx}_0$ to be  minimizer in $A_0 $ of the quadratic term  $ \frac12   \| d A_0  \|^2$ subject to the constraints
that $ \cQ A_0 = A_1$ and that $A_0$ is in the axial gauge.      This  is a linear function of $A_1$ and is written
$ A^{\min,\sx} _0 =  H^{\sx} _0 A_1 $.  We now substitute  $A_0 = A^{\min,\sx} _0 + Z_0 $ and integrate over $Z_0$ rather than $A_0$.
The free gauge action splits as 
\be
 \frac12   \| d A_0  \|^2 =  \frac12   \| d A^{\min,\sx}_0  \|^2+  \frac12   \| d Z_0  \|^2
\ee
The delta functions in (\ref{polly2}) become $  \de (\cQ Z_0)  \de_{\sx} (Z_0)$.   We identify a Gaussian measure $d \mu_{C_0} $ with covariance $C_0$
and a normalization factor  $\fZ_0$ by 
\be \label{song} 
\fZ_0 \  d \mu_{C_0}  ( Z_0 ) =   \   \de (\cQ Z_0)  \de_{\sx} (Z_0)  \exp\B( -    \frac12   \| d Z_0  \|^2 \B)  DZ_0
\ee 
The conditions  $QZ_0 = 0$ and $Z_0$ axial imply $\| d Z_0 \|^2  \geq \const  \| Z_0 \|^2$ so this is non-degenerate. 
The exact meaning of (\ref{song})  requires parametrizing the hyplane selected by the delta function  \cite{Bal85b}, \cite{Dim15b}. 

Now  the expression (\ref{polly2})  becomes 
\be       \label{polly3} 
\tilde  \rho_1(A_1,  \Psi_1  )   =  \fZ_0  \exp( -   \frac12   \| d A^{\min,\sx} _0  \|^2 )    \int     \rho^f_1\B (A^{\min,\sx}_0 + Z_0 ,\Psi_1\B) ) d \mu_{C_0} ( Z_0 ) 
\ee
The integral over $Z_0$ is a fluctuation integral,  that is an integral over the deviations from the minimizer.

Before proceeding we change gauges in the minimizer.    Instead of the minimizer   $A^{\min,\sx} _0$      in   axial gauge we consider  the 
minimizer   $A^{\min} _0$   in the Landau gauge.  It is defined to be the minimizer of  $ \frac12   \| d A_0  \|^2$ subject to the constraints
that $ \cQ A_0 = A_1$ and a version of the condition  that  the divergence of $A_0$ vanishes.    It has the form $A^{ \min}_0  = H_0 A_1$.    These two minimizers are gauge equivalent:  $ A^{\min,\sx} _0=
A^{\min} _0 + d \om $.   Thus in gauge invariant positions like  $ \| d A^{\min,\sx} _0  \|^2$  and $ \rho^f_1\B (A^{\min,\sx}_0 + Z_0 ,\Psi_1\B) ) $
we can replace $A^{\min,\sx} _0$ by $A^{\min} _0$.    The advantage of the Landau gauge is that we have better control over the minimizing operator $H_0$.
In particular one can establish that it has an exponentially decaying kernel and that it  has good bounds  on low order derivatives.   The remark about derivatives is not so important in this first step where we are on a unit lattice.  But after many steps the minimizers will be on increasingly fine lattices and control on derivatives is important.

The Landau gauge  minimizer  $A^{ \min}_0 $ is only a function of $A_1$ and we introduce the more suggestive notation $\tilde \cA_1 =  A^{ \min}_0 = H_0 A_1$.
So now we have
     \be       \label{poly4} 
\tilde   \rho_1(A_1,  \Psi_1  )   =\fZ_0  \exp\B( -   \frac12   \|d  \cA  \|^2 \B)    \int     \rho^f_1\B ( \cA  + Z_0 ,\Psi_1\B) ) d \mu_{C_0}  ( Z_0 )\ \B|_{\cA= \tilde \cA_1} 
\ee
Note that the fluctuation field   stays in axial gauge which is better for convergence of the integrals.

Next in   $  \rho^f_1(\cA  + Z_0 ,\Psi_1) )$ we have a term  $<  \bar \Psi_0,   \fD_{e_0}  (  \cA  + Z_0        )\Psi_0>$
and we write
\be  \label{bingo} 
<  \bar \Psi_0,   \fD_{e_0}  ( \cA + Z_0     ) \Psi_0>=<  \bar \Psi_0,   \fD_{e_0}  (  \cA       )\Psi_0> + V_0 ( \cA, Z_0, \Psi_0) 
\ee 
Then $V_0$ depends on $Z_0$ through 
\be
e^{ ie_0 ( \cA + Z_0 ) }   -  e^{  i  e_0  \cA  }   = e^{  i e_0   \cA  }  (  e^{i e_0   Z_0 } -1 ) 
\ee
Introducing a large/small field split in $Z_0$  (thanks to $d\mu_{C_0}(Z_0)$)    one can arrange  $|Z_0| \leq p(e_0) $ and so 
$|  e^{i e_0   Z_0 } -1 | \leq \cO(e_0 p(e_0) )$  with the same bound for the kernel of  $V_0$.  So $V_0$ is very small.    Similar remarks apply to the term 
  $ <   \bPsi_1 -  Q(- \cA -Z_0 ) \bPsi_0,  \Psi_{1} -  Q(\cA +Z_0 ) \Psi_k >$;   subtracting off the expression at $Z_0 =0$ gives another small term which
  we include in the definition of $V_0$.

 Now we have
 \be  \label{edward} 
\begin{split}
 &\rho^f_1( \cA  + Z_0,\Psi_1)\\
 & = \sN_0
 \int     D \Psi_0   \exp\B( - \tilde  \fS_1 ( \cA , \Psi_1, \Psi_0) 
 -    V_0 (  \cA , Z_0, \Psi_0)    - m_0  <  \bar \Psi_0,   \Psi_0 > 
-     \vep_0   | \bbT^0_N|         \B) \\
\end{split} 
\ee
Here we have isolated the quadratic form  
\be
\begin{split}
&\tilde  \fS_1 ( \cA , \Psi_1, \Psi_0)  \equiv  bL^{-1}   <   \bPsi_1 -  Q(-\cA  ) \bPsi_0,  \Psi_{1} -  Q(\cA  ) \Psi_0 >+  <  \bar \Psi_0,   (\fD_{e_0}   (\cA )  + \bar    m_0 ) \Psi_0 >   \\
\end{split}
\ee
Let  $\Psi^{\crit}_0(\cA ) $ be the critical  point of this form  in $\Psi_0 $. 
 and similarly define $\bPsi^{\crit}_0(\cA)  $. 
Explicitly we have 
\be
\Psi^{\crit}_0(\cA ) =  H_0(\cA ) \Psi_1  =  bL^{-1} \  \Ga_0 (\cA ) Q^T(-\cA )  \Psi_1
\ee
where 
\be \label{gamma} 
 \Ga_0(\cA )  =   \B(\fD_{e_0}  (\cA )  + \bar    m_0  +bL^{-1}\  Q^T(-\cA  ) Q(\cA ) \B)^{-1} 
\ee
There is a similar expression for $\bPsi^{\crit}_0(\cA ) $.

We  expand around the critical point. 
With the  more suggestive notation
$\tilde  \psi_1 ( \cA) = \Psi^{\crit}_0(\cA ) = H_0(\cA ) \Psi_1$
we substitute  $\Psi_0 =\tilde  \psi_1 ( \cA) + W_0$  and integrate over 
$W_0 $ rather than $\Psi_0 $. 
The quadratic form splits as
\be
\tilde \fS_1 ( \cA , \Psi_1,  \tilde  \psi_1 ( \cA)  + W_0) =\tilde  \fS_1(\cA , \Psi_1, \tilde  \psi_1 ( \cA) )   +   <  \bW_0,   \Ga_0(\cA )^{-1}  W_0 > 
\ee
We  also  identity  a Grassmann Gaussian integral  with  covariance $ \Ga_0(\cA )$ and  formal measure   $ d \mu_{\Ga_0(\cA )}(W_0)  $ given by 
\be
 \fZ_0(\cA )\   d \mu_{\Ga_0(\cA ) } (W_0 )   =     \exp \B( -<  \bW_0,   \Ga_0(\cA )^{-1}  W_0 > \B) DW_0. 
\ee
where  $ \fZ_0(\cA  )  = \det (\Ga_0(\cA )^{-1}   )$. 
  The interaction  is now  $ V_0 ( \cA , Z_0, \tilde  \psi_1 ( \cA) + W_0)  $.
   We  absorb into it the difference between the mass  counterterm  at   $ \tilde  \psi_1 ( \cA) + W_0$ and at   $\tilde  \psi_1 ( \cA)$, and call the
result $ V_0 ( \cA , Z_0,  \tilde  \psi_1 ( \cA),  W_0) $.   
So now (\ref{edward}) has become
\be  \label{edward2} 
\begin{split}
 \rho^f_1(\tilde \cA_1+ Z_0,\Psi_1)
& = \sN_0 \fZ_0( \cA)\  \exp \B( -\tilde  \fS_1(\cA , \Psi_1,   \psi )   - m_0  <     \bpsi  ,    \psi  > 
-      \vep_0     | \bbT^0_N|      \B)  \\
&
\int    d \mu_{\Ga_0(\cA ) } (W_0 )  \exp  \B(    V_0 ( \cA , Z_0,   \psi,  W_0 )   
  \B)  \B|_{\cA = \tilde \cA_1,\psi= \tilde \psi_1(\tilde \cA_1)}  \\
 \end{split} 
\ee

We add some comments about the operators $\Ga_0(\cA), H_0( \cA)$.  Controlling the inverse in the definition (\ref{gamma})  of $\Ga_0(\cA)$
is difficult unless one has a bound on  the field strength $ d\cA$.    Thanks to the factor  $e^{-\frac12 \| d \cA \|^2} $ in (\ref{poly4}) we can
supply  such a bound with a large/small field split.  In addition  it is useful to establish that the  kernels  of $\Ga_0(\cA), H_0(\cA)$ 
have exponential decay.     The mass $\bar m_0$ in    $ \Ga_0(\cA)  $ is too tiny  to get any useful estimate,  but the term  $bL^{-1}\  Q^T(-\cA ) Q(\cA)$
provides an effective mass of order $\cO(L^{-1})$ to do the job for  $ \Ga_0(\cA)  $ and hence  for $H_0(\cA)$. \cite{BOS91},\cite{Dim15b}  
The  Landau gauge minimizer $H_0$ also has exponential decay. \cite{Bal88a}, \cite{Dim15b}.  Since  $V_0 ( \tilde \cA_1 , Z_0,   \tilde \psi_1(\tilde \cA_1),  W_0 )$ has a local structure,   some approximate locality in the fundamental fields $A_1,\Psi_1$ is preserved.  
 
 We insert the fermion integral (\ref{edward2}) back into the full density (\ref{poly4})  and find 
  \be 
\begin{split}
\tilde  \rho_1(A_1,  \Psi_1  ) &
  =\sN_0 \fZ_0 \fZ_0( \cA)\  \Xi _0 \B( \cA,  \psi \B) \\
  & \exp \B( -   \frac12   \| d  \cA  \|^2-\tilde  \fS_1( \cA, \Psi_1,   \psi )  - m_0  <   \bpsi ,   \psi   > 
-      \vep_0   | \bbT^0_N|      \B)   \B|_{\cA = \tilde \cA_1, \psi= \tilde \psi_1(\tilde \cA_1) } 
 \\
 \end{split} 
\ee
where 
\be \label{cluster1} 
\Xi _0 \B(  \cA,    \psi\B)=
\int    d \mu_{\Ga_0(\cA) } (W_0 )    d \mu_{C_0}  ( Z_0 ) \chi_0 ( Z_0 ) \exp  \B(    V_0 (  \cA, Z_0,   \psi,  W_0 )    \B)
\ee
is the complete fluctuation integral.   Here we have now made explicit a    characteristic function $\chi_0 ( Z_0 )$ enforcing $|Z_0| \leq p(e_0) $
 
We need to write the fluctuation integral as the exponential of  an expression with good local properties.    First we need norms on elements of the Grassman algebra.   Let $\{ \Psi(x) \} $ be the generators of a Grassman algebra 
indexed by a unit lattice.  A general element has the form 
\be
 E(\Psi) = \sum_{n=0}^{\infty}  \frac{1}{n!}\   \sum_{x_1, \dots,  x_n}  E_n(x_1, \dots , x_n )  \Psi (x_1) \cdots  \Psi (x_n) 
 \ee
  Then  for $h>0$ a   norm is defined  by 
 \be
 \| E \|_h = \sum_{n=0}^{\infty}  \frac{h^n }{n!}  \sum_{x_1, \dots,  x_n}   |E_n(x_1, \dots , x_n )|  
 \ee
This generalizes to families of generators like our  $\{\bPsi_{0,\al} (x), \Psi_{0, \al} (x) \} $ as well as  fields defined on lattices with spacing $\ep$ in which 
in which case the sum over $x_i$ would be weighted by $\ep^3$ approximating integrals. 

Next  we write the interaction $V_0$  in a form that can reproduce  itself. 
 These are polymer expansions.   Divide the torus $\bbT^0_{N} $ into cubes $\sq$ of  width $M = L^m$ much larger than $L$.  
A polymer is defined to be a connected union of $M$-cubes.  A polymer expansion is a function of the form $f =  \sum_X f(X) $ where the sum is over polymers $X$ and $f(X) $ only depends on fields
in $X$.   We generally assume also that $f(X)$ is exponentially decaying in the size of $X$ in the sense that  $\| f(X) \|_h= \cO (  e^{-\ka d_M(X) } )$
for a sufficiently large constant $\ka$.   The quantity $d_M(X)$ is  defined by
\be
  Md_M(X)   = \textrm{ length of the shortest continuum tree joining the cubes in  }  X  
\ee 
and   the  factor   $e^{- \ka d_M(X) } $ gives tree decay on the scale $M$.  This is sufficient for the convergence of the sum over $X$.

We  express   $V_0$  as polymer functions  by 
\be
V_0 = \sum_{\sq} E_0 (\sq)  = \sum_X E_0(X) 
\ee
where first  we break the integrals up into integrals $E_0(\sq)$  over $M$-cubes $\sq$  and  then we
define $E_0(X)$ to be $E_0(\sq)$ if $X =\sq$ and zero otherwise.   
We have
$ \| E_0(\sq)\|_h  \leq \cO(e_0 p(e_0) )$ and then trivially   
\be
 \| E_0(X, \cA, Z_0)\|_h  \leq \cO(e_0 p(e_0) ) e^{- \ka d_M(X) } 
 \ee   

The evaluation of the fluctuation  integral  (\ref{cluster1})  uses a standard technique  known as the  cluster expansion.  (See for example \cite{Dim13a}).    This says that  there exist polymer
functions $ E^{\#} _0 (X)$ such that
\be
\begin{split}
&\int    d \mu_{\Ga_0( \cA) } (W_0 )    d \mu_{C_0}  ( Z_0 ) \chi_0 ( Z_0 ) \exp  \B(  \sum_{X}    E_0 (X,   \cA, Z_0,   \psi,  W_0 )  \B)  \\
&= \exp \B( \sum_X   E^{\#} _0 (X,   \cA,    \psi\B) \\
\end{split} 
\ee
With our   restrictions on the size of $d  \cA$     and for  some   $\ka' < \ka$   they also   satisfy a bound  
\be \label{sugar} 
\|  E^{\#} _0 (X, \cA )\|_h  \leq   \cO(e_0 p(e_0) )e^{- \ka' d_M(X) } 
\ee
The key ingredients in the cluster expansion  are that the   initial   functions $E_0(X)$ are sufficiently small, true by our assumption that $e_0p(e_0) $ is small,  and the exponential decay of the covariances $C_0, \Ga_0(A_0)$
of the Gaussian measures.    In carrying out the proof it is useful to  use the identities  
\be
\begin{split} 
\int f(Z_0)\   d \mu_{ C_0} (Z_0)   =  &    \int f\B(C_0^{\frac12}  Z_0\B)  d \mu_{I } (Z_0)   \\
\int  f(\bar W_0, W_0 )\  d \mu_{\Ga_0( \cA)  } ( \bar W_0, W_0  )   =  & \int  f\B( \bar W _0, \Ga_0( \cA)  W_0 \B) d \mu_{I  } ( \bar W_0, W_0  ) \\
\end{split}
\ee
which put the non-locality in the function rather than the measure.  This compromises  the polymer expansion since 
$ E_0 (X,   \cA, C_0^{\frac12} Z_0,   \psi, \bar W_0,  \Ga_0(\cA) W_0 )  $ now depends on $Z_0,W_0$ outside of 
$X$.    But one can restore it by a further decomposition using   
random walk expansions for  $C_0^{\frac12}$ and $\Ga_0( \cA) $.  With a new polymer expansion in the exponent  and   the ultra-local measures $ d \mu_{I }(Z_0) $ and  $d \mu_{I  } ( \bar W_0, W_0 )$
the cluster expansion is just    a combinatoric problem. 

With bounds on $d  \cA$ one can also establish a polymer expansion for the normalizing factor  $ \fZ_0(\cA  )  = \det \B( \fD_{e_0}  (\cA )  + \bar    m_0  +bL^{-1}\  Q^T(-\cA  ) Q(\cA )    \B)$ of the form   \cite{Dim15b}
\be
\fZ_0(\cA) = \fZ_0(0 ) \exp \B( \sum_X   E^{\det} _0 (X,   \cA) \B)
\ee
and $ E^{\det} _0 (X,   \cA) $ has a bound  similar to (\ref{sugar}).

Now  we have  with  $ E_0^{\#} = \sum_X  E_0^{\#}(X) $, etc.
 \be  \label{edward3} 
\begin{split}
\tilde  \rho_1(A_1,  \Psi_1  ) 
 & =N_0 \fZ_0 \fZ_0(0 )
   \exp \B( -   \frac12   \| d  \cA  \|^2- \tilde \fS_1( \cA, \Psi_1,  \psi  )\\
   & -m_0  <  \bpsi  ,   \psi  > 
-    \vep_0  | \bbT^0_N|     +    E_0^{\#}(   \cA,  \psi )   +E_0 ^{\det} ( \cA)  \B)    \B|_{\cA = \tilde \cA_1,  \psi= \tilde \psi_1(\tilde \cA_1)} 
 \\
 \end{split} 
\ee

Next we  scale this expression   defining  $\rho_1( A_1, \Psi_1)= \tilde \rho_1( A_{1,L} , \Psi_{1,L}) $  for $A_1, \Psi_1$ on $\bbT^0_{N-1} $.
Scaled fields   $\cA_1, \psi_1(\cA_1) $  on the $L^{-1}$ lattice  $\bbT^{- 1}_{N-1} $  are given by  
$\cA_{1,L}  =   H_0 A_{1,L }  $ and
$[ \psi_1 ( \cA_1) ] _L   =  H_0( \cA_{1,L}  )  \Psi_{1,L}    $.
Thanks to our choice of scaling factors the     $\| d    \cA \|^2 $  is invariant and   the  quadratic fermion term  becomes
\be \label{hilly} 
\begin{split} 
 \fS_1 (\cA_1,  \Psi_1, \psi_1(\cA_1) )&   \equiv 
 b<   \bPsi_{1} -  Q(-\cA_1  ) \bpsi_1(\cA_1) ,  \Psi_{1} -  Q(\cA_1  ) \psi_1(\cA_1)  > \\
& +  <   \bpsi_1(\cA_1) ,   \B(\fD_{e_1}  (\cA_1)  + \bar    m_1 \B)  \psi_1(\cA_1)  > \\
\end{split} 
\ee
Parameters    change  by 
\be \label{onward1} 
e_1 = L^{\frac12} e_0 \hs \bar m_1 = L \bar m_0  \hs  m_1 = L m_0  \hs  \bar \vep_1= L^3 \vep_0
\ee
and we have up to a multiplicative constant
 \be  \label{edward4} 
\begin{split}
  \rho_1(A_1,  \Psi_1  ) 
  =&
   \exp \B( -   \frac12   \| d  \cA_1  \|^2-   \fS_1( \cA_1, \Psi_1,  \psi_1 ( \cA_1)) \\
   & -m_1  <  \bpsi_1 ( \cA_1) , \psi_1 ( \cA_1) > 
-    \vep_1  | \bbT^0_{N-1}|     +    E_1(   \cA_1, \psi_1( \cA_1) )   \B) 
 \\
 \end{split} 
\ee
The function $ E_1(   \cA, \psi ) $ is defined as follows.  It  has a polymer expansion $ E_1  = \sum_X  E_1(X) $ where
\be \label{sparrow2} 
E_1(X,  \cA, \psi)  =   ( \cL   E_0^{\#} )(X,  \cA, \psi) +( \cL E_0  ^{\det} ) (X,  \cA, \psi) 
\ee 
Here the linear operator $\cL$ reblocks and scales.   It is defined for an $M$-polymer $X$ by
\be  \label{sparrow1} 
( \cL  E) (X, \cA, \psi  )  =  \sum_{Y: \bar Y = LX} E(Y, \cA_{L},  \psi _L   )
\ee
The sum is over all $M$-polymers $Y$  such that $\bar Y =LX$, where $\bar Y $ is the union of all $LM$ polymers intersecting $Y$.

The sum in (\ref{sparrow1}) tends to increase norms by a factor $L^3$.   This is not a problem in this first step,  but  is a key issue   to be dealt with when 
we repeat the procedure.  The remedy will involve modifying the simple scaling  in  (\ref{onward1}).

\section{Iteration}

Now we repeat the first step and  generate the sequence of densities  $\rho_1, \rho_1, \rho_2, \dots$.  
With  $A_k,\Psi_k$ on  $\bbT^0_{N-k}$    
we pass   from $\rho_k$ \ to $\rho_{k+ 1}$  by block averaging
\be   
\begin{split}     \label{polly} 
&\tilde \rho_{k+1}(A_{k+1} ,  \Psi_{k+1}  )   =  \sN_k
   \int  DA_k   D \Psi_k  \  \de ( A_{k+1} - \cQ A_k)  \   \de_{\sx}  (A_k)     \\
   &  \exp\B( - bL^{-1} <  \bPsi_{k+1} -  Q(- \tilde \cA_{k+1} ) \bPsi_k,  \Psi_{k+1} -  Q(\tilde \cA_{k+1} )  \Psi_k  > \B)    \rho_{k} (A_k,  \Psi_k)     \\
\end{split}    
\ee 
and then  scaling  $\rho_{k+1}(A_{k+1}, \Psi_{k+1} )   = \tilde \rho_{k+1} (A_{k+1,L}, \Psi_{k+1,L } ) $.
The  field   $\tilde \cA_{k+1}$ is yet  to be specified.

  We continue to 
consider    the case where the small field region is the whole torus.   If this is the only contribution  the claim is that  up to a multiplicative constant  
 \be \label{money} 
\begin{split} 
 \rho_k(A_k, \Psi_k  )  = &
   \exp \B( -   \frac12   \| d \cA _k  \|^2- \fS_k(\cA_k, \Psi_k, \psi_k( \cA_k)   ) \\
   & -m_k  <  \bar\psi_k( \cA_k) ,   \psi_k( \cA_k)  > 
-      \vep_k|  \bbT^0_{N-k}|    +    E_k (  \cA_k,  \psi_k( \cA_k)  )   \B) \\
 \end{split}
 \ee
Here  $\cA_k = \cH_k A_k$
 is a  $k$-step Landau gauge   minimizer.  The operator $\cH_k$ is      generated as the composition of $k$ single step operators.  Similarly the 
  $\psi_k(\cA_k) = \cH_k(\cA_k) \Psi_k$  and  $\cH_k(\cA)$ is generated as the  composition of $k$ single step operators. 
The fields $\cA_k , \psi_k(\cA_k)$     are defined on the finer lattice  $\bbT^{-k} _{N-k} $, so we are heading back toward the original problem on $\bbT^{-N}_0$, even though
the number of fundamental variables $A_k, \Psi_k$ is decreasing.     The quadratic fermion term 
is 
 \be  
 \begin{split}
 \fS_k (\cA_k,  \Psi_k, \psi_k(\cA_k) )   \equiv  \ & b_k    < \bPsi_{k } -  Q_k (-\cA_k) \bpsi_k(  \cA_k) , \Psi_{k } -  Q_k (\cA_k) \psi_k(  \cA_k)  >\\
 & +  <   \bpsi_k(\cA_k) ,   \B(\fD_{e_k}  (\cA_k)  + \bar    m_k \B)  \psi_k(\cA_k)  > \\
 \end{split} 
\ee 
The   scaled  coupling constant  is $e_k=L^{k/2} e_0 = L^{-(N-k)/2 }  e  $ and $Q_k(\cA_k)$ is a $k$-fold averaging operator.  
The function $E_k$ has a polymer expansion $E_k = \sum_X E_k(X)$.    The   $  E_k (X,  \cA_k,  \psi_k( \cA_k) )   $  are  restrictions of  functions $  E_k (X,  \cA,  \psi )   $
which  are  invariant  under lattice symmetries
 and gauge transformations.  But note that the gauge invariance only holds in positions corresponding to the smeared fields $ \cA_k,  \psi_k( \cA_k) $, not
 in the fundamental variables  $A_k, \Psi_k$.

We have discussed why  $\rho_1$ has the form (\ref{money}).   We now assume $\rho_k$ has the form (\ref{money}) and discuss   why  $\rho_{k+1}$ has this form.  
We insert (\ref{money}) into (\ref{polly}) and   follow the treatment  of the first step.    However   there are   new  features.   Let $A_k^{\min, \sx}= H^{\sx}_k A_{k+1}  $ be the minimizer in $A_k$ of the term $\| d \cA_k\|^2$ subject to the constraints $A_{k+1} = QA_k$ and $A_k$ axial.  Expand the action 
around  the minimizer by $A_k =   A_k^{\min,\sx}  + Z_k$.     Then $\cA_k   = \cH_k A_k $ becomes $ \cH_k A^{\min,\sx}_k + \cH_k Z_k $. 
The $\cH_k A^{\min, \sx}_k=\cH_k  H^{\sx}_k A_{k+1}$ is mixed Landau and axial gauge, but it is gauge equivalent to the all Landau $\tilde  \cA_{k+1} \equiv \cH_k  H_k A_{k+1} $
which will scale  to $\cA_{k+1}  = \cH_{k+1} A_{k+1}$.    We make this change for better estimates.   Also defining  $\cZ_k= \cH_k Z_k$ we can now write 
\be
\cA_k = \tilde \cA_{k+1} + \cZ_k 
\ee
and in  the action we have the split
 \be 
 \frac12   \| d \cA_k \|^2    = \frac12 \|d \tilde  \cA_{k+1} \|^2 + \frac 12 \| d \cZ_k \|^2 \ee
 The translation $\cA_k = \tilde \cA_{k+1} + \cZ_k $ induces changes   in the quadratic fermion action and now also  in the existing polymer functions.
 As   in (\ref{bingo})  we separate out a leading term and a fluctuating term by  
\be \label{contra1} 
\begin{split} 
 \fS_k\B(\tilde \cA_{k+1} + \cZ_k, \Psi_k, \psi_k( \tilde \cA_{k+1} + \cZ_k)   \B) = &  \fS_k\B(\tilde \cA_{k+1} , \Psi_k, \psi_k( \tilde \cA_{k+1} )   \B)  + [ \cdots ] \\
E_k\B( \tilde \cA_{k+1} + \cZ_k,  \psi_k( \tilde \cA_{k+1} + \cZ_k\B)   =& E_k\B(  \tilde \cA_{k+1} ,  \psi_k( \tilde \cA_{k+1} ) \B)   + [ \dots ] \\
\end{split}
\ee
The fluctuation  terms  $[\cdots ]$ will be discussed further.

The quadratic fermion term now has the form  at $\cA  =\tilde  \cA_{k+1} $
\be   \label{bonus} 
bL^{-1} <\bPsi_{k+1}- Q ( -   \cA ) \bPsi_k,    \Psi_{k+1}- Q ( \cA ) \Psi_k  > +  \fS_k (  \cA ,  \Psi_k, \psi_k( \cA) )        
\ee
Let   $\Psi^{\crit}_k(\cA)   = H_k( \cA)\Psi_{k+1}  $ be the critical point  in $\Psi_k$ of this expression. 
   We expand  around this field by 
by $\Psi_k =  \Psi^{\crit} (\cA)  + W_k$.   The field  $\psi_k(\cA) $ becomes
\be
\psi_k(\cA)  = \tilde  \psi_{k+1}  (  \cA ) + \cW_k( \cA)
\ee
where  $\tilde  \psi_{k+1}  (  \cA )=  \cH_k ( \cA)  H_k ( \cA) \Psi_{k+1} $ will scale to $ \psi_{k+1}  (  \cA )$ and  $\cW_k  ( \cA) =  \cH_k ( \cA) W_k $.
The quadratic fermion term  (\ref{bonus}) splits as
\be
 \tilde  \fS_{k+1} ( \cA,  \Psi_{k+1},\tilde  \psi_{k+1} ( \cA ) ) + < \bar W_k, \Ga_k( \cA ) ^{-1} W_k>  \\ 
\ee
Here  $\Ga_k( \cA  )$ is a certain invertible operator similar to  (\ref{gamma})  and 
\be
\begin{split} 
& \tilde  \fS_{k+1} ( \cA,  \Psi_{k+1}, \psi ) \\
&=   b_k   L^{-1}  < \bPsi_{k+1} -  Q_{k+1} (-\cA) \bpsi , \Psi_{k+1} -  Q_{k+1} (\cA) \psi  >
 +  <   \bpsi,   \B(\fD_{e_k}  (\cA)  + \bar    m_k \B)  \psi  > \\
 \end{split} 
\ee
where $Q_{k+1} (\cA)  = Q (\cA) Q_k (\cA) $.
The term  $ E_k(  \cA, \tilde  \psi_{k+1} (  \cA ) )$ in (\ref{contra1})  becomes
\be \label{contra2} 
E_k\B(   \cA,  \tilde  \psi_{k+1}  (  \cA ) + \cW_k( \cA) \B)  =  E_k\B(  \cA, \tilde  \psi_{k+1} (  \cA ) \B)   + [ \dots ] 
\ee
The fluctuation part $[ \dots ]$ will be discussed further. 

Continuing as in the first step we  identify Gaussian integrals by 
\be \label{Gauss} 
\begin{split} 
\fZ_k \  d \mu_{C_k}  ( Z_k ) =  & \   \de (\cQ Z_k)  \de_{\sx} (Z_k)  \exp\B( -    \frac12   \| d \cH_k Z_k  \|^2 \B)  DZ_k \\
 \fZ_k(\cA )   d \mu_{\Ga_k(\cA ) } (W_k )   = &     \exp \B( -<  \bW_k,   \Ga_k(\cA )^{-1}  W_k > \B) DW_k \\
\end{split}
\ee
Then we find
 \be  
\begin{split}
\tilde  \rho_{k+1}(A_{k+1},  \Psi_{k+1}  ) 
  &=\sN_k \fZ_k \fZ_k( \cA)\  \Xi _k \B( \cA,  \psi \B) \exp \B( -   \frac12   \| d  \cA  \|^2-\tilde  \fS_{k+1}( \cA, \Psi_{k+1},   \psi )  \\
  & - m_k  <   \bpsi ,   \psi   > 
-      \vep_k   | \bbT^0_{N-k}|  + E_k(\cA, \psi)    \B)   \B|_{\cA = \tilde \cA_{k+1}, \psi= \tilde \psi_{k+1}(\tilde \cA_{k+1}) } 
 \\
 \end{split} 
\ee
where  the fluctuation integral is 
\be 
\Xi _k \B(  \cA,    \psi \B)=
\int    d \mu_{\Ga_k(\cA) } (W_k )    d \mu_{C_k}  ( Z_k ) \chi_k ( Z_k ) \exp  \B(   E_k'  (  \cA, \cZ_k,   \psi,  \cW_k(\cA)   )    \B)
\ee
Here again we have implicitly  made a large/small field split to introduce a characteristic function $ \chi_k ( Z_k )$ enforcing a bound like  $|Z_k| \leq p(e_k)=(- \log e_k)^p $.

The    $E'_k$ are the   terms $ [\cdots] $ in (\ref{contra1}),(\ref{contra2}). They  inherit an expansion      $E'_k= \sum_X E' _k(X) $. 
However the          $E'_k(X, \cA, \cZ_k,   \psi,  \cW_k(\cA)  ) $  depend  on the fundamental variables   $Z_k, W_k $ outside   $X$ since 
   $\cZ_k = \cH_kZ_k$ and $\cW_k(\cA)   = \cH_k( \cA) W_k$ and  the kernels of     $\cH_k,  \cH_k( \cA) $ are
only   exponentially decaying.  When we shift to ultralocal Gaussian measures with  unit covariance it gets worse  with  $ \cZ_k = \cH_kC_k^{\frac12} Z_k $ and $\cW_k(\cA)  = \cH_k(\cA) \Ga_k (\cA)  W_k$.      The remedy as before  is to break $\cZ_k, \cW_k( \cA)$  and hence    $E'_k(X) $  into local 
pieces  using   random walk expansions, now   for  each the operators  of $\cH_k,  C _k^{\frac12} , \cH_k( \cA),  \Ga_k (\cA) $.  
This leads to   a polymer expansion with $Z_k,W_k$  strictly localized.   Then     one can do a cluster expansion as before. 
The result is that the fluctuation integral has the form 
\be
  \Xi _k( \cA, \psi )   = \exp \B(     E_k^\#  ( \cA, \psi ) \B)    = \exp\B( \sum_X E_k^\#(X, \cA, \psi )  \B)
\ee
with good bounds on $E_k^\#(X) $.

We also need a polymer expansion for $ \fZ_k( \cA) = \det ( \Ga_k(\cA)^{-1}  )$,  the Gaussian normalization factor   in (\ref{Gauss}).
  With restrictions on the size of $d \cA$   one can show
\be
 \fZ_k( \cA)  =  \fZ_k( 0)  \exp \B(   E_k^{\det}(\cA)   \B)   = \fZ_k( 0) \exp  \B(  \sum_X E_k^{\det} (X, \cA) \B) 
\ee
with good bounds on $E_k^{\det}(X) $.

 At this point  we have  the generalization of  the $k=0$ case (\ref{edward3})   
    \be  \label{edward5} 
\begin{split}
&\tilde  \rho_{k+1} (A_{k+1} ,  \Psi_{k+1}    ) =  \sN_k \fZ_k \fZ_k(0)\ 
   \exp \B( -   \frac12   \| d \cA   \|^2-  \tilde \fS_{k+1} (\cA, \Psi_{k+1},   \psi )   \\
  &   -m_k  <   \bpsi ,  \psi   >  -  \vep_k  | \bbT^0_{N-k} |  +   E_k  (  \cA ,   \psi)      +    E^{\#} _k (  \cA ,    \psi )  +  E_k^{\det} (  \cA)   \B) 
  \B| _{\cA = \tilde \cA_{k+1}, \psi= \tilde \psi_{k+1}(\tilde \cA_{k+1}) } 
 \\
 \end{split} 
\ee

Before scaling we make some further adjustments. 
Our goal is to show that things do not grow too rapidly as we iterate the procedure.    There is a potential problem with the polymer functions $E_k(X) $ which 
tend to grow like $L^3$ in each step as noted at the end of section \ref{four}.   However this only occurs for terms with a few fields, since scaling
also supplies a factor $L^{-\frac12} $ for gauge fields, and $L^{-1} $ for fermion fields.  Furthermore we gain powers of $L^{-\frac12}$ from 
 $e_k=  L^{-\frac12}e_{k+1}$.     Taking account the symmetries of the $E_k(X)$  the only relevant terms
are the energy density (constants) and the fermion mass ( $\int \bpsi \psi$  ).  Boson mass terms  $\int |\cA|^2$ are forbidden by gauge invariance.   Other
terms shrink and are said to be irrelevant;  for example four fermion terms shrink like $L^{-1}$.    There are no marginal terms.

    Accordingly we can write 
\be
E_k(X, \cA, \psi )   = -\vep_k^*(X)  \Vol(X)   -   \int \bpsi  [m_k^*(X) ] \psi   + (\cR E_k) (X,\cA, \psi )  
 \ee
and with a suitable choice of  $\vep_k^*(X) , m_k^*(X) $ the remainder $(\cR E_k) (X)$  shrinks  under scaling.  
Summing over $X$ gives
 \be
\sum_X  E_k(X,\cA, \psi )   =  -\vep^*_k   | \bbT^0_{N-k} |   -  m^*_k \int \bpsi    \psi   + \sum_X (\cR E_k) (X,\cA, \psi )  
 \ee
where  $\vep^*_k  = \sum_{X \supset \sq} \vep_k ^*(X) $ and    $m^*_k = \sum_{X \supset \sq} m^*_k (X) $ for any $M$-cube $\sq$.

We make these adjustments in (\ref{edward5}).  
After scaling  we find that up to a constant  $\rho_{k+1}$ has the claimed form:   
 \be \label{money2} 
\begin{split} 
& \rho_{k+1}(A_{k+1}, \Psi_{k+1}  )  = 
   \exp \B( -   \frac12   \| d \cA \|^2- \fS_{k+1}(\cA, \Psi_{k+1}, \psi  ) \\
 & \hs \hs    -m_{k+1}  <  \bar\psi,   \psi > 
-     \vep_{k+1}|  \bbT^0_{N-{k+1}}|    -    E_{k+1} (  \cA,  \psi    )  \B) \B| _{\cA =  \cA_{k+1}, \psi= \psi_{k+1}( \cA_{k+1}) } 
 \\
 \end{split}
 \ee
 The coupling constant in $\fS_{k+1}$ is now   $e_{k+1} =   L^{\frac12 }e_k $ and we have the  new parameters: 
 \begin{equation}  \label{flow1} 
 \begin{split} 
\vep_{k+1} = &  L^3  (\vep_k  + \vep^* _k )\\
   m_{k+1} = &  L  (m_k + m^*_k  )\\
E_{k+1} =&   \cL  (  \cR  E_k   + E_k^\#  + E_k^{\det} )      \\
\end{split} 
\end{equation} 
Here $\vep^*_k= \vep^*_k(E_k)$, $m^*_k=  m^*_k(E_k)$,    $E_k^\#=E_k^\#(m_k,E_k ) $,  and  $\cL $ is the  reblocking and scaling operation (\ref{sparrow1}). 
  The possibility of dangerous growth is now isolated in $m_k, \vep_k$. 
 
 We need to study the flow of these equations,  a problem in discrete dynamical systems. 
 We start with $E_0 =0$ and  the renormalization problem is to choose initial values  $\vep_0 = \vep^N_0$  
$m_0 = m ^N_0$   such that   $\vep_k$ and $m_k$ take specified final values and $E_k$ stays small. 
We stop after $K<N $ steps with  $N-K = \cO(1) $,  and  arbitrarily ask for final values   $\vep_K=0, m_K=0  $.  (This does not give massless fermions since the background
mass is   $\bar m_K  = L^{-(N-K) } \bar m >0.)$    The existence of a unique solution can be formulated as a fixed point problem in a Banach space of  sequences
   $\{ \vep_k, m_k, E_k \}^K_{k=0} $ satisfying the boundary conditions.   The fixed point exists if $e$ is  sufficiently small.  
This leads to a  result that looks something like the following \cite{Dim15b}.

\begin{thm} \label{pasta}    Let $m>0$,  and  $e$ be sufficiently small.   Then for any $N$ and   $N-K = \cO(1)$  there exists a unique sequence  $\{ \vep_k, m_k, E_k \}^K_{k=0} $ satisying   (\ref{flow1}) with the initial condition $E_0 =0$ and the final condition $\vep_K =0, m_K =0$.    With   $e_k = e^{-(N-k))/2} e$ and   $h_k = e_k^{-\frac14}$,  the solution satisfies
\be \label{flow2} 
|\vep_k|    \leq  e_k^{\frac14}    \hs     |m_k|   \leq e_k^{\frac34}   \hs  \|E_k(X)\|_{h_k}  \leq  e_k^{\frac14}   e^{-\ka d_M(X) }
\ee
With  this sequence the density $\rho_k$ has the form (\ref{money}) for all  $0 \leq k \leq K$. 
\end{thm} 
\bigskip

The fractional exponents here are an artifact of the proof and have no fundamental significance.

Theorem \ref{pasta}  gives good bounds for the  partition function $\sZ_N(e)$    if one takes it to be  $  \int  \de_{\sx} (A_K) \rho_K(A_K  ,\Psi_K ) DA_k D \Psi_K $.
But  this  density $\rho_K$  is only  the global small field  contribution.    For the true partition function we have to include the contribution from  terms which have some
 large field regions.

The  history of small field regions is   given by a non-increasing  sequence  
 $\bom= ( \Om_1, \cdots, \Om_k) $ with   $\Om_j$ a union of $L^{-(k-j)} M$ cubes in $\bbT^{-k}_{N-k} $.  It is created as follows.
    In passing from $k$ to $k+1$  a new small field region   $\Om_{k+1}$  is created.  It  is a union of $LM$ cubes (or larger)  in $\bbT^{-k}_{N-k}$,
and   is defined with bounds  depending on $p(e_{k+1} )= (-\log e_{k+1} )^p$
 to keep pace  with the running coupling constant $e_{k+1}$.  The  $p(e_k)$ are decreasing in $k$ so the small field constraints are becoming tighter as we
 proceed.      The new small field regions are  
introduced    only where needed, namely  inside the old small  field region:    $\Om_{k+1} \subset \Om_k$.   At the end of  step $k+1$ the  $\Om_{k+1}$ is scaled down to a union of $M$ cubes in  $\bbT^{-k-1}_{N-k-1}$.

The complete expression for  $\rho_k $ involves a sum over  histories   $\bom$:
\be
\rho_k(A_k  ,\Psi_k )  = \sum_{\bom}  \rho_{k, \bom}(A_k  ,\Psi_k )  
\ee
In  the small field region $\Om_k$ for  $ \rho_{k, \bom}$  the action density is identical with the global small field case we have discussed at length.
The contribution of the large field region $\Om_k^c$ is much smaller  is estimated along the lines discussed in section \ref{split}.

 There are difficulties with carrying out this program. 
 The density  $ \rho_{k, \bom}(A_k,\Psi_k )  $ is  naturally expressed in terms of smeared fields 
 $\cA_{k, \bom},  \psi_{k,\bom}$ which depend on the history $\bom$.   These   depend on the fundamental fields $A_k,\Psi_k $
 through operators   
$\cH_{k, \bom} $ and  $\cH_{ k, \bom}(\cA )$.  Estimates and local expansions for these operators are needed and
the tool is again random walk expansions.      But because the history $\bom$ has a multiscale structure   these are multi-scale random walk expansions.  
These  operators  also  embody       coupling between large and small field regions,  which is abundant and needs to be controlled.  
     Another area of difficulty is the presence of  localized 
characteristic functions  which can  be a source of non-locality when subjected to expansions around critical points. 
Finally at each step there is not a single large/small split as our notation would indicate,   but  several such.  Things get rather complicated.

These difficulties can be overcome,
and  when we divide by the free partition function we  get the stability    bound of Theorem   \ref{one}, which we recall is the bound uniform in $N$
\be
 K_- \leq \sZ_N(e)/\sZ_N(0) \leq K_+
\ee

\end{document}